\documentclass[aps,prl,twocolumn,groupedaddress,showpacs]{revtex4-1}
\usepackage{graphicx}
\usepackage{hyperref}

\begin{document}


\title{Frequency Multiplexed Magnetometry via Compressive Sensing}


\author{Graciana Puentes$^{1,2}$, Gerald Waldherr$^{1}$, Philipp Neumann$^{1}$, and J\"{o}rg Wrachtrup$^{1}$}

\affiliation{\noindent 1-III Physikalisches Institut, Research Center SCOPE, and MPI for Solid State Research, University of Stuttgart, Pfaffenwaldring 57, 70569 Stuttgart, Germany. \\
\noindent 2-ICFO - The Institute of Photonic Sciences, Mediterranean Technology Park, Av. Carl Friedrich Gauss 3, 08860 Castelldefels, Barcelona, Spain}

\date{\today}

\begin{abstract}
\noindent  
Quantum sensors based on single Nitrogen-Vacancy (NV) defects in diamond are state-of-the-art tools for nano-scale magnetometry with precision scaling inversely with total measurement time $\sigma_{B} \propto 1/T$ (Heisenberg scaling) rather than as the inverse of the square root  of $T$, with   $\sigma_{B} = 1/\sqrt{T}$ the Shot-Noise limit. This scaling can be achieved by means of phase estimation algorithms (PEAs) using adaptive or non-adaptive feedback, in combination with single-shot readout techniques. Despite their accuracy, the range of applicability of PEAs is limited to periodic signals involving single frequencies with negligible temporal fluctuations. In this Letter, we propose an alternative method for precision magnetometry in frequency multiplexed signals via compressive sensing (CS) techniques. We show that CS can provide for precision scaling approximately as $ \sigma_{B} \approx 1/T$, both in the case of single frequency and frequency multiplexed signals, as well as for a 5-fold increase in sensitivity over dynamic-range gain, in addition to reducing the total number of resources required.

\end{abstract}
\maketitle


\noindent Toward the realisation of highly sensitive magnetic field sensors operating at room temperature and with atomic resolution, nano-scale magnetometry experiments in solids have been performed using single nitrogen-vacancy centres (NV) in diamond (Fig. 1 (a)), achieving detection of very weak magnetic fields ($B \approx 3$ nT) with spatial resolution of a few nanometers \cite{Maze, Balasubramanian}. However, standard magnetometry precision is limited by statistical fluctuations, the so-called Shot-Noise limit $\sigma_{B} =1/ \sqrt{T}$ \cite{Taylor}, where $T$ is the total time required to estimate the magnetic field. This scaling is due to the fact that in a standard measurement $n=T/\tau$ independent measurements are performed over a short time-interval $\tau$, yielding a magnetic field precision $\sigma_{B} \approx 1/\sqrt{\tau T}$. Therefore, it should in principle be  possible to achieve precision scaling as $\sigma_{B} \approx 1/T$, if one were to perform a single measurement over the entire period ($\tau=T$). This is the maximum precision possible for a phase measurement, and is referred to as the Heisenberg limit \cite{Caves,Yurke,Berry2}.

Notwidthstanding, there are at least two problems hindering Heisenberg-limited precision in solid state magnetometry. The first is  spin relaxation which precludes measurements longer than the dephasing time. The second is that performing measurements over long periods usually results in ambiguities. A solution to eliminate the phase-ambiguity problem is the implementation of quantum phase estimation algorithms (QPEAs) \cite{Cleve, NielsenChuang}. QPEAs are based on applying the inverse Quantum Fourier Transform (QFT) \cite{Shor}, which can be implemented using local measurements and control. The problem with just using QPEA is that it produces a probabilty distribution with large tails, with precision far from the Heisenberg limit. This additional problem can in turn be overcome by applying feedback schemes to achieve $1/T$ scaling \cite{Higgins}. A remarkable approach to achieving Heisenberg-like precision scaling based on a phase estimation algorithm (PEA) without adaptive feedback, was proposed in Ref. \cite{Berry}, and successfully implemented in Ref. \cite{Waldherr}, in combination with single-shot read-out techniques \cite{Neumann}. In order to prevent ambiguities, the maximum magnetic field range is  [$-\Delta B_{max}, \Delta B_{max}$), which limits the longest accumulation time to:

\begin{equation}
\tau_{0} < \frac{\pi}{2 \gamma \Delta B_{max}},
\end{equation}

where the phase accumulation during a Larmor precession can be expressed $\phi(\tau)=\omega \tau$, with $\omega=\gamma B$ the Zeeman shift (Fig. 1 (b)). PEAs have been applied to several other problems of interest, such as reference-frame alignment \cite{Chiribella}, clock synchronization \cite{Joza}, frequency \cite{Rosenband} and position measurements \cite{Taylor}, in addition to electric field sensing \cite{Dolde}. However, a basic conditions for the implementation of PEAs is the assumption that the signal is composed of a \emph{single} frequency. This can be a significant limiting factor in the presence of temporal fluctuations, or for frequency multiplexed signals where more than one characteristic Larmor frequency may be involved. In such situation, the standard approach is to repeat $n$ independent Ramsey measurements at equally distributed times ($n \tau_{0}$) up to the dephasing time, thus scaling as $1/\sqrt{T}$.\\

\begin{figure}[t!]
\hspace{-0.7cm}
\includegraphics[width=0.50\textwidth]{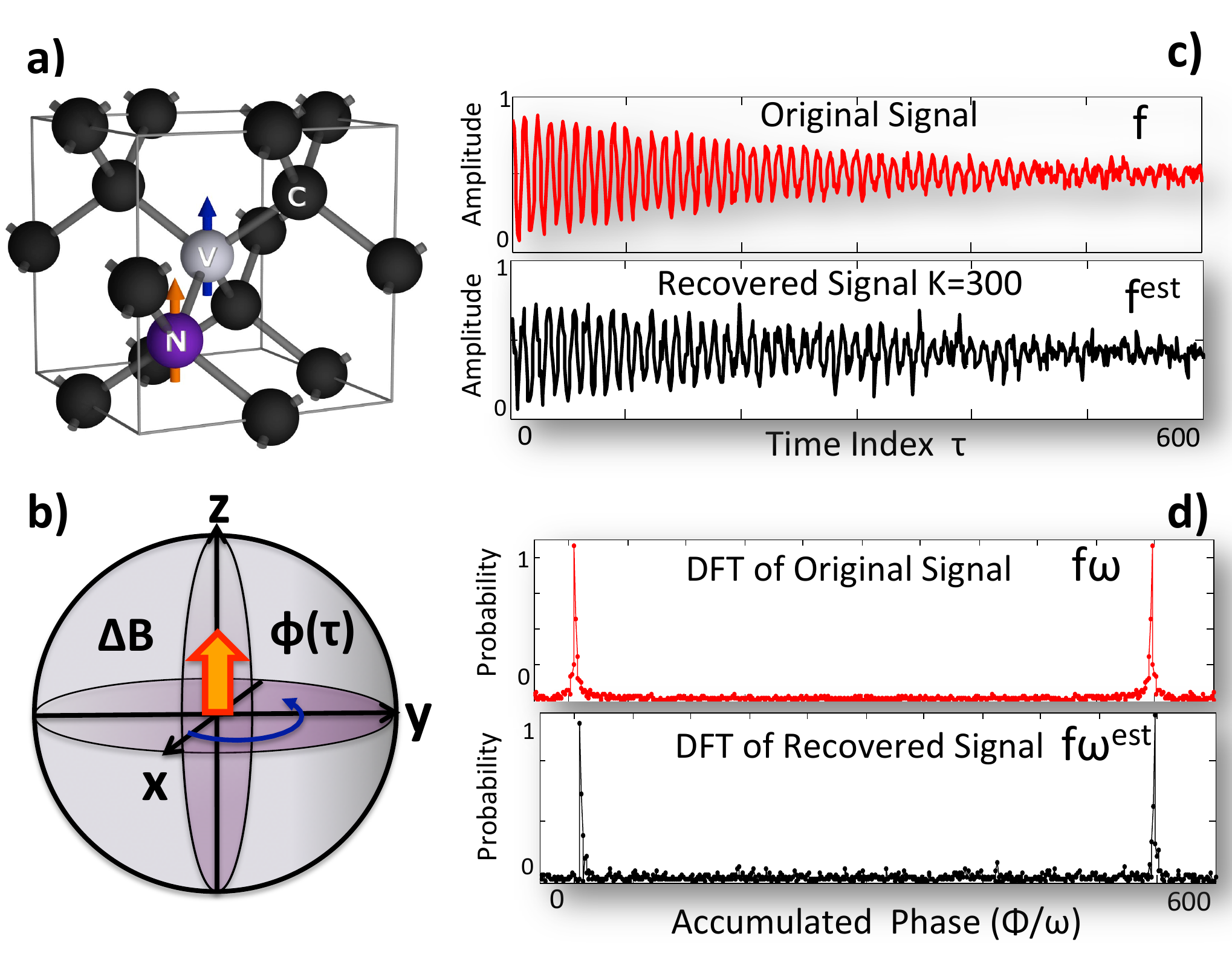}
 \caption{(a) Scheme of NV structure in diamond lattice. (b) Bloch sphere illustration of Larmor precession $\phi(\tau) =\omega \tau$ around magnetic field $\Delta B$. (c) Top: Original signal given by Larmor precession typically obtained by Ramsey interferometry \cite{Waldherr}. The total time  $N=600$ determines the dimension of the basis for compressed sensing; (c) Bottom: Recovered signal by compressive sensing after applying a measurement operator $A_{(K \times N)}$ using a subset of $K=N/2$ random data points with a uniform distribution. (d) Top: Discrete Fourier Transform (DFT) of original signal displaying the frequency sparsity of the input. The appearance of a second peak located at $N-\omega$ is a numerical effect due to the periodicity of the DFT; (d) Bottom: Recovered signal via compressive sensing in frequency domain.}
\end{figure}

Motivated by this relevant yet complex scenario, we present an alternative approach for multi-frequency magnetometry involving compressive sensing (CS) techniques. CS algorithms are extensively employed in the context of signal processing to recover sparse vectors from a reduced number of measurements. Here sparsity refers to a few non-zero components in a given N-dimensional basis (Fig. 1 (d) top), and the measurement constraints, defined by a suitable measurement operator $A$, are linear functions of the inputs. When the measurements are chosen at random, the original signal (Fig. 1 (c) top) can be uniquely determined from a small number of measurements ($K<N$) via efficient convex optimization routines (Fig. 1 (c) bottom, and 1 (d) bottom). CS is therefore a highly suitable tool for magnetometry - a sparse problem in frequency - since it satisfies all the above mentioned criteria. CS algorithms have been readily successfully applied to computational biology \cite{Dei} and graphics \cite{Sen}, medical imaging \cite{Lustig}, communication theory \cite{Griffin}, in addition to quantum state tomography \cite{Gross} and quantum process tomography \cite{White} of fairly pure density matrices and low rank quantum operators. In this Letter, we show that compressive sensing reconstructions can provide for Heisenberg-like precision scaling, both for the case of single-frequency and multi-frequency magnetometry, thus extending their range of applicability beyond the scope of PEAs, in addition to providing for a 5-fold increase in sensitivity over bandwidth gain as compared with standard measurements. Moreover, we show CS can reduce the total number of resources subject to the complexity of the input signal. \\

\noindent \emph{Single-Frequency Magnetometry via CS.-} In order to compare the performance of CS with PEAs in the case of single-frequency magnetometry we consider an analogous input signal ($\vec{f}(\tau)$) to the one previously analysed by Waldherr \emph{et al.} \cite{Waldherr} (Fig. 1 (c) Top). We transfrom this signal to the frequency domain by means of a Discrete Fourier Transform (DFT) algorithm obtaining $\vec{f_{\omega}}$ (Fig. 1 (d) Top), in order to perform random spectral data sampling. We fix the spectral resolution to $\Delta \omega=1/N$ throughout the search, so that all the points in the frequency domain contain information about the full input signal, eventhough the spectral resolution could be modified adaptively, as discussed below. The maximal detectable frequency is set to $1/\tau_{0}$, with $\tau_{0}$ given by the upper bound in Eq. (1). The number of sampling points in the frequency domain for the CS algorithms are increased exponentially in the form $n_{k}=n_{0} 2^{k}$, where $k=0,1,...,K$ ($K=10$), with  $n_0=N/2^k$, resulting in experimental result vectors $\vec{w}^{exp}_{k}$, with $n_{k}$ independent elements following Ref. \cite{Waldherr}. The compressive sensing algorithm is implemented by way of defining $A_{(k,N)}$ ($k=0,..,K$) measurement operators consisting of $n_{k}$ random raws of the identity matrix ($I_{N \times N}$), and searching for the most probable vector $\vec{f}_{w}^{est}$, which satisfies the measurement constraints $A_{(k,N)} \vec{f_{w}^{est}}=\vec{w}^{exp}_{k}$. Since $A_{(k,N)}$ is not a square matrix for $n_{k}<N$, it is non-invertible and the set of linear constraints is underdetermined. The key to the reconstruction is to impose non-linear regularization involving $l1$-norm minimization \cite{Kosout,Gross2}. The search can thus be casted into a convex optimization problem of the form:

\begin{equation}
\mathrm{minimize} \hspace{0.2cm} || \vec{f}_{w}^{est}||_{1}, \hspace{0.5cm} \mathrm{s.t.} \hspace{0.5cm} A_{(k,N)} \vec{f}_{w}^{est}=\vec{w}^{exp}_{k},
\end{equation}

\begin{figure}[t!]
\hspace{-0.5cm}
\includegraphics[width=0.50\textwidth]{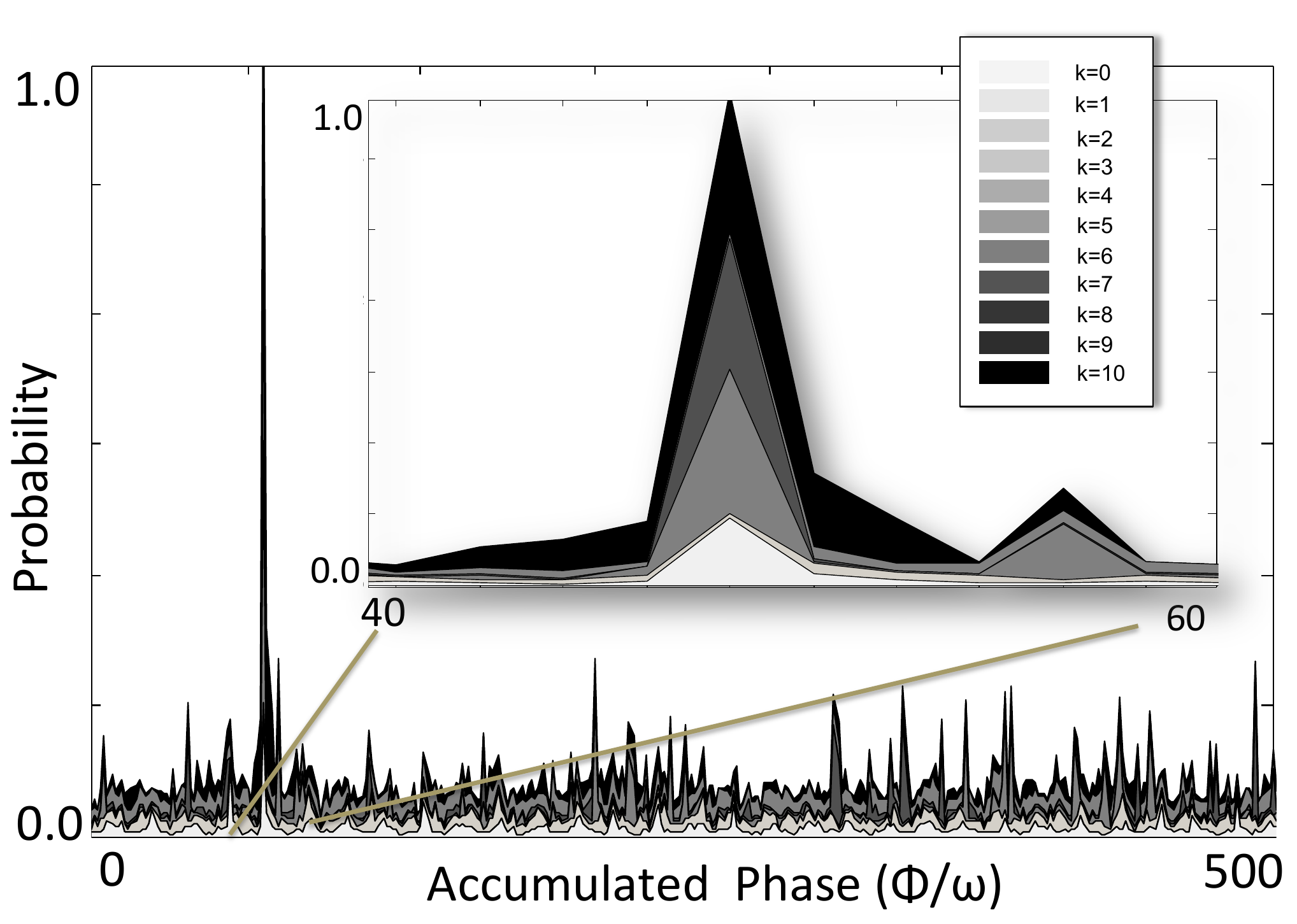}
 \caption{Typical probability distribution for the estimated phase $\phi$ via compressive sensing techniques. Probabilities in different grey tones correspond to increased number of random sample points given by $n_{k}= n_{0}2^{k}$, with $k=0,1,...,10$. Since the phase search is casted into a convex optimization problem there is no ambiguity (i.e., no local maxima) for a sufficient number of measurements.  }
\end{figure}

where we choose a flat distribution as the initial guess ($  \vec{f_0}_{w}^{est}$). Numerical results for the output of the convex search are shown in Fig. 2. The  phase $ \phi(\tau)=\omega \tau$ is normalized to the Zeeman shift ($\omega$) and is proportional to the time index ($\tau$). Since the phase search is casted into a convex optimization problem, a clear unambiguous peak (i.e., no local maxima) is present even for small number of random sampling points, thus confirming that CS can reduce resource requirements. \\

\begin{figure}[b!]
\hspace{-0.3cm}
\includegraphics[width=0.5\textwidth]{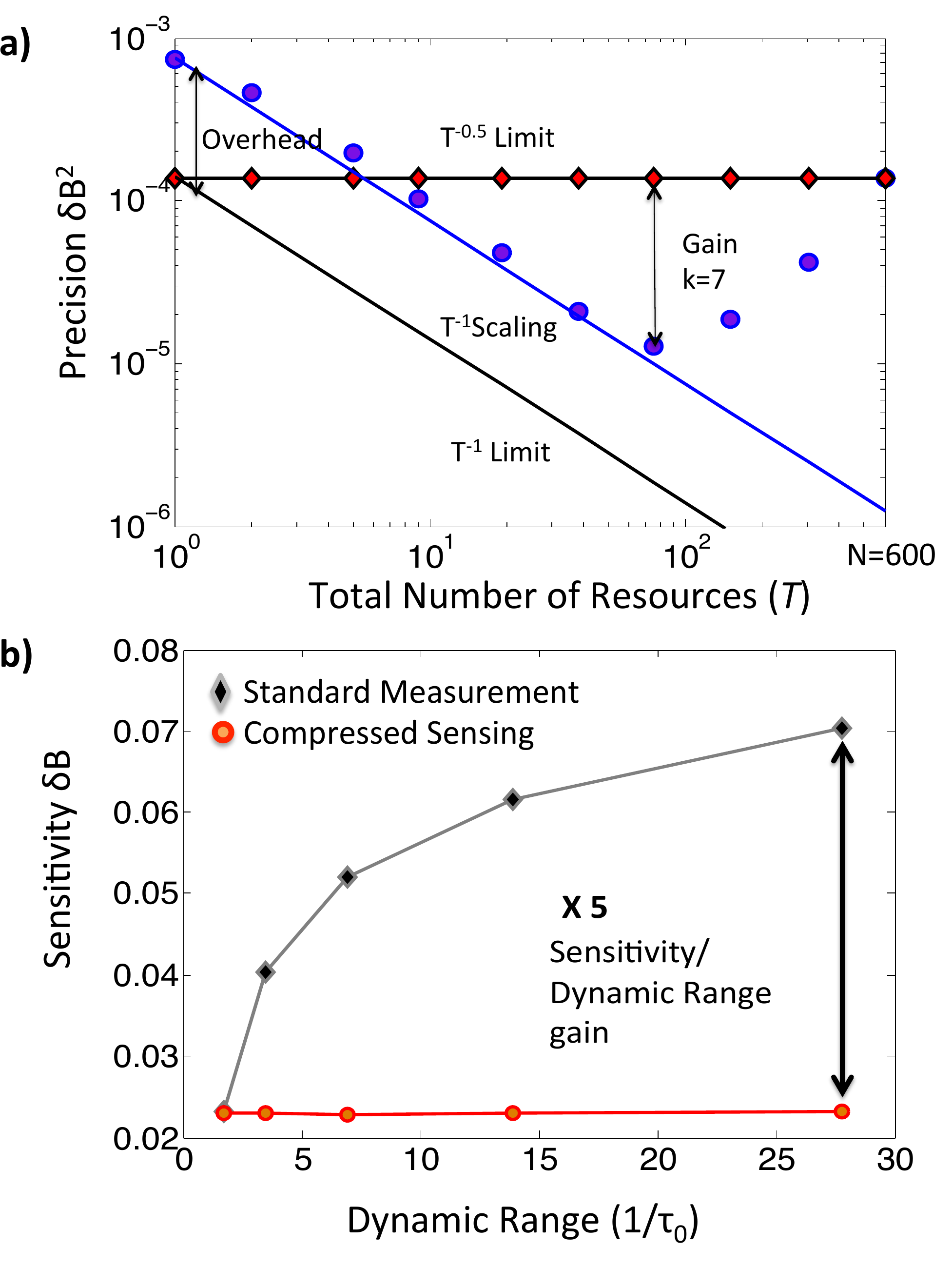}
 \caption{(a) Precision scaling ($\sigma_{B}^2 T$) of phase estimation via compressive sensing (CS) vs. total time resources ($T$). The black line indicates the Heisenberg limit ($1/T$). Red dots indicate the Shot-Noise limit ($1/\sqrt{T}$) given by the standard measurement. The numerical data is fitted to $\approx 1/T$ (blue line) showing that CS can provide for Heisenberg-like scaling. The overhead is set by the tolerance in the $l1$-norm minimization. Precision gain over standard measurement using compressive sensing  is maximal for $k=7$. (b) Dependence of sensitivity ($\delta B$) vs. maximal frequency over magnetic field $(1/\tau_{0})$ for the standard measurement (black rombos) as compared to CS approach (red dots). The area between the curves indicates the gain in sensitivity over dynamic range, showing a 5-fold increase with respect to standard measurements.}
\end{figure}

We are interested in the scaling of the variance ($\sigma_{B}^2$) of the estimated vector on the total measurement time ($T$), corresponding to the square of the magnetic field sensitivity defined as $\delta B= \sqrt{\sigma_{B}^2 T}$ \cite{Waldherr}. Where the variance is defined as the square of the norm-2 between the estimated vector ($\vec{f}_{w}^{est}$) and the vector of experimental results ($\vec{w}^{exp}_{k}$), i.e., $\sigma_{B}=|| A\vec{f}_{w}^{est} - \vec{w}^{exp}_{k}||_{2}$. Numerical results for the scaling of precision ($\sigma_{B}^2 T$) vs. total number of resources ($T$), and for the dependence of sensitivity $\delta B= \sqrt{\sigma_{B}^2 T}$ vs. dynamic range $1/\tau_{0}$ are displayed in Fig. 3. Figure 3 (a) shows the Heisenberg limit (black line) and a fit to the numerical points obtained via CS, revealing Heisenberg-like scaling $\approx 1/T$ for phase estimation via CS techniques. Red dots correspond to the Shot-Noise limit ($1/T^{0.5}$), set by the variance in the standard measurment. CS gives a maximumm precision gain as compared to  the standard measurement for $k=7$, which represents less than 16 $\%$ of the total resources ($N=600$). The number of resources required for a maximum precision gain could be further reduced by means of adaptive techniques, as discussed below. The overhead in Fig. 3 (a) is determined by the tolerance of the $l1$-norm minimization. We note that the absolute precision yielded by the CS inversion is somewhat arbitrary, we are therefore interested in the \emph{scaling} of the precision with the total number of resources. Figure 3 (b) displays the sensitivity ($\delta B$) of the recovered phase via CS (red dots), for different maximum frequency over magnetic field values (i.e., dynamic range $1/\tau_{0}$) obtained by increasing $\Delta B_{max}$, such that $\tau_{0}=0.036,0.072, 0.144, 0.288, 0.576$. The selected range of $\tau_{0}$ permits to increase the dynamic range by 4 bits (from $K=10$ to $K=14$). This is compared with the standard measurement scheme (black rombos) showing a 5-fold increase in sensitivity over dynamic range gain, via CS data recovery. The results presented in Fig. 2 and Fig. 3 confirm that CS techniques can provide for a similar performance as compared with PEAs \cite{Waldherr}, for a reduced number of resources (i.e., total measurement data points) for the case of single-frequency signals. Below we present an application of CS techniques in situations where PEAs are not applicable.  \\

\begin{figure}
\hspace{-0.8cm}
\includegraphics[width=0.52\textwidth]{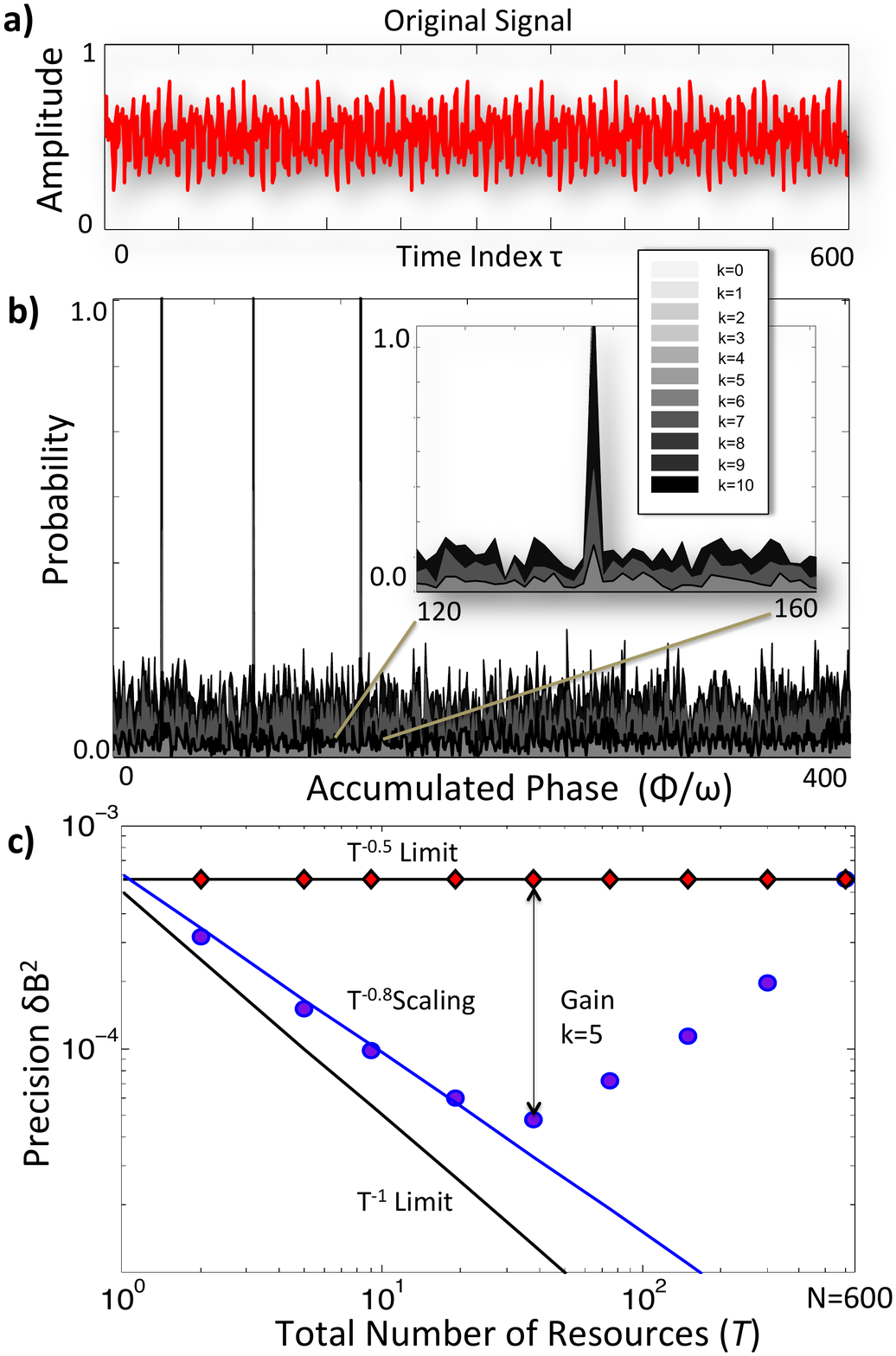}
 \caption{(a) Simulated frequency multiplexed signal consisting of Larmor precessions given by 3 arbitrary Zeeman shifts $\omega_{1,2,3}$ and $N=600$ points. (b) Typical probability distribution for multiple phase estimation, different tones correspond to increased number of random sample points given by $ n_{k}= n_{0}2^{k}$, with $k=0,1,...,10$. (c) Precision scaling ($\sigma_{B}^2 T$) vs. total time resources ($T$). The precision using CS scales as $\approx 1/T^{0.8}$ (blue line). The Heisenberg limit  ($1/T$) is indicated with a black solid line. Red dots correspond to the Shot-Noise limit ($1/T^{0.5}$) set by the standard measurement. Maximum precision gain via CS reconstruction is obtained for $k=5$.} 
\end{figure}

\noindent \emph{Frequency-Multiplexed Magnetometry via CS.-} We now present the main result of the paper which consists of applying CS recovery techniques to frequency multiplexed signals. Frequency multiplexed magnetometry can arise, for instance, when the NV sensor is coupled to a spin bath, so that hyperfine coupling results in different Zeeman splittings and different frequencies for Larmor precessions, thus resulting in effective nutations. This scenario is clearly prohibitive for the use of PEAs, however it is clearly amenable for the use of CS approaches. We consider an input signal given by the sum of three Fourier components with arbitrary  Zeeman splittings $ \omega_{1}$, $\omega_{2}$ and $\omega_{3}$ of the form $f(\tau)= \sin( \frac{2\pi  \omega_{1}}{N}\tau)+\sin( \frac{2\pi  \omega_{2}}{N}\tau)+\sin( \frac{2\pi  \omega_{3}}{N}\tau)$, with $N=600$ (Fig. 4 (a)). In this expample, we consider the same amplitude and off-set phase for the three Fourier components, although this is not a necessary requirement. Indeed, CS can also be used to determine the unknown amplitude relation and off-set phase of the  components. We increase the number of random sample data points in the same exponential manner $n_{k}=n_{0} 2^{k}$ with $k=0,1,...,10$, keeping $(1/\tau_{0})$ fixed. The recovered phase for different values of $k$ is shown in Fig. 4 (b), displaying three unambigous peaks, corresponding to the multiple Larmor frequencies $\omega_{1,2,3}$ even for small number of random sampling points, thus confirming that CS can reduce resource requirements also for signals consisting of multiple frequencies. 

Figure 4 (c) shows the precision scaling ($\sigma_{B}^2 T$) vs. total time resources ($T$) for CS data inversion in the case  of frequency multiplexed signals. The numerical data points are fit to $\approx 1/T^{0.8}$, with a precision overhead given by the tolerance in the $l1$-norm minimization algorithm. The Heisenberg limit is indicated with a black line. Red dots correspond to the Shot-Noise limit ($1/T^{0.5}$), set by the variance in the standard measurement. Eventhough the precision scaling is slightly lower in the frequency-multiplexed case, the overhead is also reduced under the assumption of ideal measurements, showing that CS techniques can outperform standard measurements even for the smallest number of resources. Furthermore, CS gives a maximum precision gain as compared to  the standard measurement for $k=5$, which represents less than $10\%$ of the total resources ($N=600$). The number of resources required for a maximum precision gain could be further reduced by means of adaptive techniques, as discussed below.  \\

We reported on a novel approach to frequency multiplexed magnetometry via compressive sensing (CS) techniques. We numerically showed that CS data recovery can provide for Heisenberg-like precision scaling ($\approx 1/T$) in situations where phase estimation algorithms (PEAs) are not applicable, in addition to providing for a 5-fold increase in sensitivity over  dynamic-range gain, and a reduction in the total time-resource requirements, subject to the complexity of the input signal. In this realisation we considered a fixed spectral resolution $\Delta \omega=1/N$ throughout the convex search, such that each point in the frequency domain contains information about the full signal in the time domain. As a future step we will consider hybrid approaches, by increasing adaptively the spectral resolution $\Delta \omega_{k}$, starting the search with a broad distribution ($\Delta \omega_{k=0}>1/N$), and narrowing down the frequency window iteratively, while increasing the number of sample points ($k$).  Other relevant hybrid approaches could be implemented by combining structured random measurement operators \cite{Gross}. Our results pave the way for potential applications of CS  in efficient quantum parameter estimation, quantum sensing, and magnetometry involving time-varying \cite{Magesan}, and frequency multiplexed signals. \\

The authors gratefully acknowledge Gopalakrishnan Balasubramanian and Friedmann Reinhard for fruitfull discussions. We acknowledge financial support by the Max-Planck-society, the EU (Squtec), Darpa (Quasar), BMBF (CHIST-ERA) and contract research of the Baden-W\"{u}rttemberg foundation. 
{}

\end{document}